\documentclass[aps,prd,twocolumn,draft,nofootinbib]{revtex4-1}
\usepackage{amsmath}
\usepackage{amsthm}

\newtheorem*{conj}{Conjecture}

\newcommand{\vnab}{\vec{\nabla}}

\begin{document}

\title{Constraints and degrees of freedom in Lorentz-violating field
  theories}

\author{Michael D. Seifert}
\affiliation{Dept.\ of Physics, Astronomy, and Geophysics, Connecticut College}
\email{mseifer1@conncoll.edu}

\date{\today}

\begin{abstract}
 Many current models which ``violate Lorentz symmetry'' do
  so via a vector or tensor field which takes on a vacuum expectation
  value, thereby spontaneously breaking the underlying Lorentz
  symmetry of the Lagrangian.  To obtain a tensor field with this
  behavior, one can posit a smooth potential for this field, in which
  case it would be expected to lie near the minimum of its potential.
  Alternately, one can enforce a non-zero tensor value via a Lagrange
  multiplier.  The present work explores the relationship between
  these two types of theories in the case of vector models.  In
  particular, the na\"ive expectation that a Lagrange multiplier
  ``kills off'' one degree of freedom via its constraint does not
  necessarily hold for vector models that already contain primary
  constraints.  It is shown that a Lagrange multiplier can only reduce
  the degrees of freedom of a model if the field-space function
  defining the vacuum manifold commutes with the primary
  constraints. 
\end{abstract}

\maketitle

\section{Introduction}

Many classical field theories are constructed in such a way that the
``most natural'' solutions to the equations of motion involve a
non-zero field value.  This paradigm, where an underlying symmetry of
the Lagrangian or Hamiltonian is spontaneously broken by the solutions
of the equations of motion, has proven to be both compelling and
fruitful over the years.  In the context of particle physics, the
best-known example is the Higgs field \cite{Englert1964, Higgs1964,
  Guralnik1964, Higgs1966}; in the context of condensed-matter
physics, this paradigm underlies the modern theory of phase
transitions, most notably the Ginzburg-Landau theory of
superconductivity \cite{Ginzburg1950}.  Many non-linear sigma models
can also be thought of in this way, if one view the model's target
manifold as being embedded in some higher-dimensional space in which
the fields are forced to a non-zero value.  Such models have been used
in the study of both particle physics \cite{Gell-Mann1960, Gursey1960}
and in ferromagnetism \cite{Joyce1967}.

In more recent years, this paradigm has also been used to study
possible observational signatures of Lorentz symmetry violation in the
context of modified gravity theories.  Such models include a new
vector or tensor field, and have equations of motion that are
satisfied when the metric is flat and the new tensor field is constant
but non-zero.  In this sense, Lorentz symmetry is spontaneously broken
in these theories, as the tensor field takes on a vacuum expectation
value that has non-trivial transformation properties under the Lorentz
group.  The vacuum state of such a model is often said to be
``Lorentz-violating'', though it would be more accurate to say that
Lorentz symmetry is spontaneously broken in the model.  Examples
of such models can be found in \cite{Kostelecky1989, Jacobson2001,
  Bekenstein2004, Kostelecky2004, Moffat2006, Altschul2010}, as well
as a particularly early example in \cite{Nambu1968}.

In general, all of these models (Lorentz-violating or otherwise) have
the property that the fields, collectively denoted by $\psi^\alpha$,
will satisfy an equation of the form $f(\psi^{\alpha}) = 0$ in the
``vacuum'', appropriately defined.  Here, $f$ is a real-valued
function of the fields; thus, if the fields $\psi^\alpha$ are
specified by $N$ real numbers, the equation $f(\psi^{\alpha}) = 0$
will generically specify an $(N-1)$-dimensional hypersurface in field
space.  I will therefore refer to this function as the \emph{vacuum
  manifold function}.

Two broad classes of models in which some fields have non-trivial
background values can readily be conceived of.  In one class, the
constraint $f(\psi^{\alpha})= 0$ is enforced exactly via the
introduction of a Lagrange multiplier $\lambda$ in the Lagrange
density:
\begin{equation} \label{LLagrange}
\mathcal{L}_{LM} = (\nabla \psi^{\alpha})(\nabla \psi^{\alpha}) +
\lambda f(\psi^{\alpha}), 
\end{equation}
where we have denoted the kinetic terms for the fields $\psi^{\alpha}$
schematically.  I will call such models \emph{Lagrange-multiplier (LM)
  models}.

In the other class, the fields $\psi^{\alpha}$ are assigned a
potential energy that is minimized when the field is non-zero.  In
particular, if we define a potential $V(\psi^{\alpha}) \propto
f^2(\psi^{\alpha})$, and write down a Lagrange density
\begin{equation}  \label{Lpotential}
\mathcal{L}_P = (\nabla \psi^{\alpha})(\nabla \psi^{\alpha}) -
V(\psi^{\alpha}), 
\end{equation}
then the lowest-energy state of the theory would be expected to occur
when $f(\psi^{\alpha}) = 0$ and $\nabla \psi^{\alpha} = 0$.  I will
call such models \emph{potential models}.

A natural question then arises: for a given collection of fields
$\psi^{\alpha}$ and a given kinetic term
$(\nabla \psi^{\alpha}) (\nabla \psi^{\alpha})$, what is the
relationship between the models \eqref{LLagrange} and
\eqref{Lpotential}?  In particular, one might expect the following
statement to be true:
\begin{conj}
  In a potential model such as \eqref{Lpotential}, the fields can take
  on any value in the $N$-dimensional field space.  By contrast, in a
  Lagrange-multiplier model, the fields are constrained to an
  $(N-1)$-dimensional subspace of field space.  Thus, the number of
  degrees of freedom of a Lagrange-multiplier model \eqref{LLagrange}
  should be one fewer than the degrees of freedom of the corresponding
  potential model \eqref{Lpotential}.
\end{conj}
The main purpose of this article is to show that this na\"ive
conjecture not is true in general;
if $\psi^{\alpha}$ includes a spacetime vector or tensor field, it may
be false.  In such models, the fields may need to satisfy certain
constraints due to the structure of the kinetic terms; adding a new
``constraint'' to such theories, in the form of a Lagrange multiplier,
does not automatically reduce the number of degrees of freedom of the
theory.

To demonstrate this, I will analyse the degrees of freedom of two
types of symmetry-breaking models.  After a brief description of
Dirac-Bergmann analysis in Section \ref{db.sec}, I will first analyse
a multiplet of Lorentz scalar fields with an internal symmetry,
followed by a vector field (Sections \ref{scalar.sec} and
\ref{vector.sec} respectively.)  For the vector fields, the analysis
will depend on the structure of the kinetic term chosen, and so three
distinct sub-cases will need to be treated.  In each case, I will
examine a model where the fields are assigned a potential energy, and
one where the fields are directly constrained via a Lagrange
multiplier.

The bulk of the explicit analysis in this work will be done in the
context of a fixed, flat background spacetime; however, I will briefly
discuss these models in the context of a dynamical curved spacetime in
Section \ref{disc.sec}. In that section, I will also discuss the
implications of these results for the broader relationship between
these two classes of models.  More general tensor fields will be
examined in a forthcoming work \cite{tensorpaper2018}.

Throughout this work, I will use units in which $\hbar = c = 1$; the
metric signature will be $({}-{}+{}+{}+{})$.  Roman indices $a, b, c,
\dots$ will be used to denote spacetime tensor indices; $i, j, k,
\dots$ will be used to denote spatial indices, where necessary.  Greek
indices $\alpha, \beta, \gamma, ...$ will be used exclusively to
denote indices in field space.  All expressions involving repeated
indices (either tensor indices or field space indices) can be assumed
to obey the Einstein summation convention.

\section{Dirac-Bergmann analysis \label{db.sec}}

Our primary tool for finding the number of degrees of each model will
be Dirac-Bergmann constraint analysis \cite{Dirac1964}; my methods and
nomenclature below will draw heavily from the later work of Isenberg
\& Nester \cite{Isenberg1977}.  I will briefly summarize the method
here, and then illustrate it in more detail via the example theories
described in Section \ref{scalar.sec}.

The method of Dirac-Bergmann analysis involves the construction of a
Hamiltonian which generates the time-evolution of the system.  In the
process of this construction, one may need to introduce constraints
among various variables, thereby reducing the number of degrees of
freedom of the system.  One may also discover that the evolution of
certain field combinations is undetermined by the equations of motion
(for example, gauge degrees of freedom).  These combinations of
fields, which we will collectively call ``gauges'', must be
interpreted as unphysical, again reducing the number of physical
degrees of freedom of the model.

If, as is usual, we count a field degree of freedom as a pair of
real-valued functions (e.g., a field value and its conjugate momentum)
that can be freely specified on an initial data surface, then the
number of degrees of freedom $N_{dof}$ can be inferred quite simply
once the above analysis is complete:
\begin{multline} 
  \label{generalcount}
  N_{dof} = \frac{1}{2} \left[ \genfrac(){0pt}{0}{\text{no.\
        of}}{\text{fields}} +
    {\text{no.\ of} \choose \text{momenta}} \right. \\
  \left. {} - {\text{no.\ of} \choose \text{constraints}} -
    {\text{no.\ of} \choose \text{gauges}} \right].
\end{multline}
In general, of course, the number of fields and the number of
conjugate momenta will be the same.  Moreover, in the particular field
theories I will be considering in this work, I will not find any
``gauges'', so the last term in \eqref{generalcount} will vanish.
Thus, for my purposes, the above equation reduces to
\begin{equation}
  N_{dof} = {\text{no.\ of} \choose \text{fields}} - \frac{1}{2}  {\text{no.\ of} \choose
    \text{constraints}}.   
\end{equation}

\section{Scalar multiplet fields} \label{scalar.sec}

The first case we will consider is a multiplet of $N$ Lorentz scalars:
$\psi^{\alpha} = \phi^\alpha$, with $\alpha = 1, 2, 3, \dots, N$.  We
wish to construct a model where these scalars ``naturally'' take on
values in some $(N-1)$-dimensional hypersurface, defined by
$f(\phi^\alpha) = 0$ (with $f$ a real-valued function.)  The
``potential model'' for this field will be derived from the Lagrange density
\begin{align} \label{scalarpot} \mathcal{L} &= - \frac{1}{2}
  \partial_a \phi^{\alpha} \partial^a \phi^{\alpha} -
  \kappa f(\phi^{\alpha})^2 \\&= \frac{1}{2} \left[
    (\dot{\phi}^{\alpha}) (\dot{\phi}^{\alpha}) - (\vec{\nabla}
    \phi^{\alpha}) \cdot (\vec{\nabla} \phi^{\alpha}) \right] -
  \kappa f(\phi^{\alpha})^2 \notag ,
\end{align}
where $\kappa$ is a proportionality constant;  the LM model for this
field will be
\begin{align} \label{scalarLM} \mathcal{L} &= - \frac{1}{2}
  \partial_a \phi^{\alpha} \partial^a \phi^{\alpha} -
  \lambda f(\phi^{\alpha}) \\&= \frac{1}{2} \left[
    (\dot{\phi}^{\alpha}) (\dot{\phi}^{\alpha}) - (\vec{\nabla}
    \phi^{\alpha}) \cdot (\vec{\nabla} \phi^{\alpha}) \right] -
  \lambda f(\phi^{\alpha}) \notag,
\end{align}
where $\lambda$ is a Lagrange multiplier field.  In both cases, we
have chosen a time coordinate $t$ and performed a 3+1 decomposition of
the tensors; a dot over a quantity (e.g., $\dot{\phi}^{\alpha}$) will
denote its derivative with respect to this time coordinate, while
spatial derivatives will be denoted with either $\vec{\nabla}$ or
$\partial_i$ depending on the expression.

\subsection{Potential model}
The degrees of freedom for the model \eqref{scalarpot} are
particularly easy to count.  The momenta conjugate to the fields are
all well-defined:
\begin{equation} \label{scalarmom}
  \pi^{\alpha} = \frac{\partial \mathcal{L}}{\partial \dot{\phi}^{\alpha}} =
  \dot{\phi}^{\alpha} 
\end{equation}
The Hamiltonian density is therefore
\begin{equation} \label{scalarpotham}
\mathcal{H}_0 = \pi^{\alpha} \dot{\phi}^{\alpha} - \mathcal{L} =
\frac{1}{2} \left[ \pi^\alpha \pi^\alpha + (\vec{\nabla}
    \phi^{\alpha}) \cdot (\vec{\nabla} \phi^{\alpha}) \right] + \kappa f(\phi^\alpha)^2.
\end{equation}
Nothing further is required here; we have no primary constraints on
the initial data, and the Hamiltonian obtained by integrating
$\mathcal{H}_0$ over space will generate the field dynamics.  We thus
have the $N$ degrees of freedom one would expect.

\subsection{Lagrange-multiplier model}
Counting the degrees of freedom for the model \eqref{scalarLM}
requires a bit more effort.  As the kinetic term of \eqref{scalarLM}
is the same as that of \eqref{scalarpot}, the momenta conjugate to the
scalars $\phi^\alpha$ are again defined by \eqref{scalarmom}.  The
difficulty arises due to the Lagrange multiplier $\lambda$.  From the
perspective of the model, it is just another field; but its associated
momentum vanishes automatically:
\begin{equation}
\varpi = \frac{\partial \mathcal{L}}{\partial \dot{\lambda}} = 0 \equiv \Phi.
\end{equation}
We thus have a \emph{primary constraint}, $\Phi = 0$, on this theory.
The ``base Hamiltonian density'' $\mathcal{H}_0 = \pi^{\alpha}
\dot{\phi}^{\alpha} - \mathcal{L}$ must then be modified to obtain the
``augmented Hamiltonian density'' $\mathcal{H}_A$ by adding this
primary constraint multiplied by an auxiliary Lagrange multiplier
$u_\lambda$:\footnote{Here and throughout, we will need to distinguish
  between the ``real'' Lagrange multiplier that appears in the
  original Lagrangian and the ``auxiliary'' Lagrange multipliers that
  are used to construct a Hamiltonian for the model.  In general, we
  will only have one real Lagrange multiplier at a time, which we will
  denote with $\lambda$; auxiliary Lagrange multipliers will be
  denoted with the symbol $u$, possibly with subscripts or diacritical
  marks.  }
\begin{equation}
  \mathcal{H}_A = \mathcal{H}_0 + u_\lambda \varpi = \frac{1}{2}
  \left[ (\pi^{\alpha})^2 +  (\vec{\nabla}
    \phi^{\alpha})^2 \right] + \lambda f(\phi^\alpha) + u_\lambda \varpi
\end{equation}

We now need to ensure that this constraint is preserved by the
time-evolution of the system; in other words, we must have
$\dot{\varpi} = \{ \varpi, H_A \} = 0$, where $H_A \equiv \int
\mathcal{H}_A d^3x$.\footnote{We are playing a bit fast and loose with
  notation here; in Hamiltonian field theory, the Poisson bracket is
  only rigorously defined for a functional with a single real value,
  not for a field which is a function of space.  A more rigorous
  definition of what we mean by an expression like $\{\varpi, H_A\}$
  is given in the Appendix.  } If this Poisson bracket does not vanish
identically, this demand will yield a secondary constraint $\Psi_1 =
0$.  The demand that \emph{this} constraint be preserved may lead to
new secondary constraints $\Psi_2 = 0$, $\Psi_3 = 0$, and so forth,
which must themselves be conserved.  We will refer to the stage at
which a secondary constraint arises as its ``order''.  In other words,
if $\Psi_1$ ensures the preservation of a primary constraint, it is a
``first-order secondary constraint''; if $\Psi_2$ ensures the
preservation of $\Psi_1$, it is a ``second-order secondary
constraint''; and so on.  In this process, it may occur that the
preservation of these constraints allows us to determine the auxiliary
Lagrange multiplier $u_\lambda$ introduced above.  The process is
continued until all constraints are known to be automatically
conserved or a contradiction is reached.

With this in mind, we derive the secondary constraints for this
model.  We first have
\begin{equation}
  0 = \dot{\varpi} = \{ \varpi, H_A \} = - \frac{\delta
    H_A}{\delta \lambda} = f(\phi^\alpha). 
\end{equation}
Thus, $\Psi_1 = f(\phi^\alpha) = 0$ is a secondary constraint.  We
repeat this procedure, obtaining another secondary constraint:
\begin{equation}
  \dot{\Psi}_1 = \{ \Psi_1, H_A \} = \frac{\partial
    f(\phi^\alpha)}{\partial \phi^\beta} \frac{\delta
    H_A}{\delta \pi^\beta} = (\delta_\beta f )\pi^\beta
  \equiv \Psi_2,
\end{equation}
where we have defined $\delta_\beta f = \delta f/\delta \phi^\beta$.
(Higher derivatives will be defined similarly.)  $\Psi_2$ must
also be conserved, which leads to a third secondary constraint:
\begin{align}
  \dot{\Psi}_2 &= \{ \Psi_2, H_A \}  \notag \\
               &= - \delta_\alpha f \left[ - \nabla^2 \phi^\alpha +
                 \lambda \delta_\alpha f \right] + \pi^\alpha
                 \pi^\beta (\delta_{\alpha \beta} f) \equiv
                 \Psi_3. \label{scalar.thirdordereq} 
\end{align}

Finally, demanding that $\Psi_3$ be conserved allows us to determine
the auxiliary Lagrange multiplier $u_\lambda$; this is because
$\Psi_3$ is itself dependent on $\lambda$:
\begin{align}
  \dot{\Psi}_3 
  &= \{ \Psi_3, H_A \} \notag \\ 
  &= \frac{\partial \Psi_3 }{\partial \phi^\gamma} \frac{\delta
    H_A}{\delta \pi^\gamma} 
    - \frac{\partial \Psi_3 }{\partial \pi^\gamma} \frac{\delta
    H_A}{\delta \phi^\gamma} 
    + \frac{\partial \Psi_3 }{\partial \lambda} \frac{\delta
    H_A}{\delta \varpi}.
\end{align}
When the dust settles, we obtain
\begin{multline}
  \dot{\Psi}_3 = \pi^{\gamma} \left[ (\delta_{\alpha \gamma} f) \left(
      3 \nabla^2 \phi^\alpha - 4 \lambda \delta_\alpha f \right) +
    \nabla^2 \left( \delta_\alpha f \right)
  \right.  \\
  \left. {} + \pi^\alpha \pi^\beta (\delta_{\alpha \beta \gamma} f)
  \right] - (\delta_\alpha f) (\delta_\alpha f) u_\lambda.
\end{multline}
So long as $\delta_\alpha f \neq 0$ when $f = 0$, this allows us to
determine the previously-unknown auxiliary Lagrange multiplier
$u_\lambda$.  Thus, the process terminates here.

Having determined the Hamiltonian and its constraints, we can count
the degrees of freedom.  We have $N+1$ fields, namely the multiplet
$\phi^{\alpha}$ ($\alpha = 1, \dots, N$) and the Lagrange multiplier
$\lambda$; one primary constraint $\Phi = \varpi = 0$; and three
secondary constraints, $\{ \Psi_1, \Psi_2, \Psi_3 \} = 0$.  The single
auxiliary Lagrange multiplier is determined, which means that there
are no unphysical gauge degrees of freedom.  With $N+1$ fields and
four constraints, the number of degrees of freedom of the model
\eqref{scalarLM} is therefore
$$
N_{dof} = (N+1) - \frac{1}{2}(4) = N-1.
$$
Thus, the Lagrange-multiplier theory \eqref{scalarLM} has one less
degree of freedom than the corresponding potential theory
\eqref{scalarpot}; in this case, the na\"ive conjecture outlined in
the introduction holds true.

\section{Vector fields} \label{vector.sec}

We now consider the case of a vector field $A_a$ which spontaneously
breaks Lorentz symmetry.  We will again restrict our attention to the
case of flat spacetime.  As the motivation behind these models is
usually a spontaneous breaking of Lorentz symmetry, we want our
Lagrange density to be a Lorentz scalar, without any prior geometry
specified.

In any such Lagrange density, we can identify a set of ``kinetic''
terms $\mathcal{L}_K$ that depend on the derivatives of $A_a$.  The
most general kinetic term that we can write down which is quadratic in
the field $A_a$ is\footnote{This follows the notation of
  \cite{Jacobson2001}, with the coefficient $c_4$ from that reference
  set equal to zero.}
\begin{equation}
\mathcal{L}_k = c_1 \partial_a A_b \partial^a A^b + c_2 (\partial_a
A^a)^2 + c_3 \partial_a A_b \partial^b A^a. \label{vecgenkin}
\end{equation}
However, since
\begin{equation}
(\partial_a A^a)^2 = \partial_a A_b \partial^b A^a + \partial_a \left[
  A^a \partial_b A^b - A^b \partial_b A^a \right],
\end{equation}
we can eliminate one of $c_2$ or $c_3$ via an integration by parts.
We will therefore set $c_2 = 0$ in what follows.  The familiar
``Maxwell'' kinetic term
$$
\mathcal{L}_K = - \frac{1}{4} F_{ab} F^{ab},
$$
with $F_{ab} = 2 \partial_{[a} A_{b]}$ corresponds to $c_1 = -c_3 =
-\frac{1}{2}$.

We can now perform the usual $3+1$ decomposition of the Lagrange density,
writing $A_0$ for the $t$-component of $A_a$ and $\vec{A}$ (or $A_i$)
for its spatial components.  The kinetic term \eqref{vecgenkin} then
becomes
\begin{multline} \label{vectordecomp} \mathcal{L}_K = \frac{1}{2}
  c_{13} \left( \dot{A}_0 \right)^2 - \frac{c_1}{2} \left( \vnab A_0
  \right)^2 - \frac{c_1}{2} \dot{\vec{A}}^2 - c_3 \dot{\vec{A}}\cdot
  \vnab A_0 \\ {} + \frac{c_1}{2} (\partial_i A_j) (\partial^i A^j) +
  \frac{c_3}{2} (\partial_i A_j) (\partial^j A^i),
\end{multline}
where $c_{13} \equiv c_1 + c_3$.  The momenta conjugate to $A_0$ and
$\vec{A}$ are then
\begin{equation}
  \Pi^0 = \frac{\partial \mathcal{L}_K}{\partial \dot{A}_0} = c_{13}
  \dot{A}_0 \label{pi0def}
\end{equation}
and 
\begin{equation}
  \vec{\Pi} = \frac{\partial \mathcal{L}_K}{\partial \dot{\vec{A}}} = - c_1
  \dot{\vec{A}} - c_3 \vnab A_0. \label{pivecdef}
\end{equation}
These equations can be inverted, to find the velocities $\dot{A}_0$
and $\dot{\vec{A}}$, so long as $c_1 + c_3 \neq 0$ and $c_1 \neq 0$.
If either of these expressions vanishes, \eqref{pi0def} and
\eqref{pivecdef} will instead yield constraint equations; we will have
to handle these cases separately.  

For the potential term, meanwhile, our desire for the Lagrangian to be
a Lorentz scalar implies that the only possible form for the vacuum
manifold function is one which sets the norm of $A_a$ to some constant
$b$.  For simplicity's sake, we will therefore choose $f(A_a)$ to be
of the following form:
\begin{equation}
f(A_a) = A^a A_a - b
\end{equation} 
where $b$ is a constant.  Depending on the sign of $b$, the ``vacuum''
manifold will consist of timelike vectors ($b<0$), spacelike vectors
($b >0$), or null vectors ($b = 0$).\footnote{As spontaneous symmetry
  breaking implies a non-zero field value, one would normally exclude
  the case $b = 0$ to ensure that $A_a = 0$ is not in the vacuum
  manifold.  However, this is not necessary for the analysis which
  follows.}  Our potential model will then be
\begin{equation} \label{vectorpot} \mathcal{L}_P = \mathcal{L}_K -
  \kappa f(A_a)^2 =\mathcal{L}_K - \kappa (-A_0^2 + \vec{A}^2 -b)^2,
\end{equation}
where $\kappa$ is again a proportionality constant; the LM model will
be
\begin{equation} \label{vectorLM} \mathcal{L}_{LM} = \mathcal{L}_K -
  \lambda f(A_a)  = \mathcal{L}_K - \lambda (-A_0^2 + \vec{A}^2 -b).
\end{equation}
For compactness, I will denote the four-norm of $A_a$ as
$A^2 = -A_0^2 + \vec{A}^2$; the norm of the spatial part on its own
will always be denoted by $\vec{A}^2$.

\subsection{General case:  $c_{13} \neq 0$, $c_1 \neq
  0$ \label{genvector.sec}} 

\subsubsection{Potential model}

Performing a Legendre transform on $\mathcal{L}_P$ \eqref{vectorpot}
to obtain the Hamiltonian density, we obtain a base Hamiltonian
density of
\begin{multline} 
  \label{vecpotham:case1} 
  \mathcal{H}_B = \frac{1}{2 c_{13}} \Pi_0^2 - \frac{1}{2 c_1}
  \vec{\Pi}^2 + \frac{c_1^2 - c_3^2}{2 c_1} \left( \vnab A_0
  \right)^2 - \frac{c_3}{c_1}  \vec{\Pi} \cdot \vnab A_0 \\
  {} - \frac{c_1}{2} (\partial_i A_j) (\partial^i A^j) - \frac{c_3}{2}
  (\partial_i A_j) (\partial^j A^i) + \kappa \left(A^2
    - b \right)^2.
\end{multline}
The resulting theory has four fields ($A_0$ and $\vec{A}$) and no
constraints; so the process terminates here, and the base Hamiltonian
is the complete Hamiltonian for the model.  Counting the degrees of
freedom, we find that
\begin{equation}
N_{dof} = 4 - \frac{1}{2}(0) = 4.
\end{equation}

\subsubsection{Lagrange multiplier model \label{genvecLM.sec}}

For the Lagrange multiplier model \eqref{vectorLM}, we have a
primary constraint associated with $\lambda$:
\begin{equation}
\varpi = \frac{\partial \mathcal{L}}{\partial \dot{\lambda}} = 0.
\end{equation}
We must therefore augment the Hamiltonian with an auxiliary Lagrange
multiplier $u_\lambda$ to enforce this constraint:
\begin{align}
  \mathcal{H}_A &= \Pi^0 A_0 + \vec{\Pi} \cdot \vec{A} - \mathcal{L}_{LM}
                  + u_\lambda \varpi \notag \\
                &= \frac{1}{2 c_{13}} \Pi_0^2 - \frac{1}{2 c_1}
                  \vec{\Pi}^2 + \frac{c_1^2 - c_3^2}{2 c_1} \left( \vnab A_0
                  \right)^2 - \frac{c_3}{c_1}  \vec{\Pi} \cdot \vnab
                  A_0 \notag \\
                & \qquad \qquad {} - \frac{c_1}{2} (\partial_i A_j) (\partial^i A^j) - \frac{c_3}{2}
                  (\partial_i A_j) (\partial^j A^i) \notag \\
                & \qquad \qquad \qquad {}  + \lambda (A^2 -
                  b) + u_\lambda \varpi
\end{align}

We now find the secondary constraints required for the primary
constraint to be conserved under time-evolution.  Taking the Poisson
brackets of each constraint with the Hamiltonian in turn, we obtain
three secondary constraints:
\begin{align}
  \dot{\varpi} &= \{ \varpi, H_A \} = A_0^2 - \vec{A}^2 + b
                 \equiv \Psi_1; \\
  \dot{\Psi}_1 &= \{ \Psi_1, H_A \} \notag \\
               & = 2 \left[ \frac{1}{c_{13}}
                 A_0 \Pi^0 + \frac{1}{c_1} \vec{A} \cdot \left( \vec{\Pi} + c_3 \vnab
                 A_0 \right) \right] \equiv \Psi_2; \\
  \intertext{and}
  \dot{\Psi}_2 &= \{ \Psi_2, H_A \} = 2 \lambda \left(
                 \frac{A_0^2}{c_{13}} - \frac{\vec{A}^2}{c_1} \right)
                 + \Xi \equiv \Psi_3, \label{genvec.thirdordereq}
\end{align}
where
\begin{multline}
  \Xi \equiv \frac{1}{c_{13}^2} \left( \Pi^0 \right)^2 -
  \frac{1}{c_1^2} \left( \vec{\Pi} + c_3 \vnab A_0 \right)^2 \\ {} +
  \frac{c_3}{c_1 c_{13}} \left( \vec{A} \cdot \vnab \Pi^0 - A_0 \vnab
    \cdot \vec{\Pi} \right) + \frac{c_1 - c_3}{c_1} A_0 \nabla^2 A_0
  \\ {} - \frac{c_3}{c_1} \vec{A} \cdot \vnab \left( \vnab \cdot
    \vec{A} \right) + \vec{A} \cdot \nabla^2 \vec{A}.
\end{multline}
All three of the quantities $\Psi_1$, $\Psi_2$, and $\Psi_3$ must
vanish for the model to be consistent.

When we take the Poisson bracket of $\Psi_3$ with $H_A$, we
will obtain
\begin{align}
  \dot{\Psi}_3 &= \{\Xi,
                 H_A\} + \left\{ 2 \lambda \left( \frac{A_0^2}{c_{13}} -
                 \frac{\vec{A}^2}{c_1} \right), H_A \right\} \notag \\
               &= \{\Xi,
                 H_A\} + 2 u_\lambda \left( \frac{A_0^2}{c_{13}} -
                 \frac{\vec{A}^2}{c_1} \right) \notag \\ 
               & \qquad {} + 4 \lambda
                 \left[\frac{1}{c_{13}^2} A_0 \Pi^0 + \frac{1}{c_1^2}
                 \vec{A} \cdot \left(\vec{\Pi} + c_3 \vnab A_0 \right) \right]
                  \label{finalconstrvecLM:case1}
\end{align}
This means that for generic initial data, for which 
\begin{equation}
\frac{A_0^2}{c_{13}} \neq
\frac{\vec{A}^2}{c_1}, \label{genvec.genericID} 
\end{equation}
we can solve \eqref{finalconstrvecLM:case1} for
$u_\lambda$.\footnote{The full expression for
  $\{\Xi, H_A \}$ is complicated and not terribly
  illuminating, so we will not present it here.  However, from the
  form of $\Xi$ and $\mathcal{H}_A$, we can see that it will depend on
  $A_0$, $\vec{A}$, $\Pi^0$, $\vec{\Pi}$, and $\lambda$---but it will
  be independent of both $\varpi$ and (more importantly) $u_\lambda$.}
Thus, the auxiliary Lagrange multiplier $u_\lambda$ is determined via
the self-consistency of the theory.  We have five fields ($A_0$,
$\vec{A}$, and $\lambda$), and self-consistency generates four
constraints (one primary, three secondary); and so the total number of
degrees of freedom of this model is
\begin{equation}
N_{dof} = 5 - \frac{1}{2}(4) = 3.
\end{equation}
Again, as expected from the conjecture, we have lost one degree of
freedom to the Lagrange multiplier.

A related analysis was performed by Garfinkle, Isenberg, and
Martin-Garcia \cite{Garfinkle2012} in the case of Einstein-aether
theory.  In such models, the vector field is constrained to satisfy
$A_a A^a = b = -1$, i.e., the vector field is unit and timelike.  In
\cite{Garfinkle2012}, the time component $A_0$ of the vector field was
explicitly eliminated from the Lagrangian after performing a $3+1$
decomposition, leaving the components of $\vec{A}$ as the three
dynamical fields.  It was found in that case that the model did not
contain any extra constraints on these three dynamical fields, if
(assuming $c_2 = c_4 = 0$, as we have done here)
\begin{equation}
  c_1 \neq 0, \qquad \frac{c_3}{c_1} \leq 0.
\end{equation}
Such models were called ``safe'' by the authors of
\cite{Garfinkle2012}; such a model would be expected to contain the
three degrees of freedom present in $\vec{A}$.  Models with $c_1 = 0$
were called ``endangered'', in that they contained additional
constraints on the initial data; such models would contain fewer than
three degrees of freedom.  Finally, models with $c_1 \neq 0$ and
$c_3/c_1 > 0$ were called ``conditionally endangered'', since the
constraint structure of the equations differed at various points in
configuration space.

To connect this to the present work, we note that
\eqref{genvec.genericID} is equivalent to
\begin{equation}
c_1 A_0^2 - c_1 \vec{A}^2 - c_3 \vec{A}^2 \neq 0,
\end{equation}
or, since $c_1 \neq 0$ and $-A_0^2 + \vec{A}^2 = b$ under the
constraint,
\begin{equation}
b \neq - \frac{c_3}{c_1} \vec{A}^2.
\end{equation}
In the case where $c_3/c_1 \leq 0$ and $b < 0$, this is guaranteed to
hold, and we therefore have no additional constraints and three
degrees of freedom.  However, if $c_3/c_1 > 0$, there can be
non-generic points in configuration space where the number of
constraints changes.

\subsection{Maxwell case:  $c_{13} = 0$, $c_1 \neq 0$ \label{Maxvec.sec}} 

\subsubsection{Potential model}

We now consider a vector field $A_a$ with a ``Maxwell'' kinetic
term, for which $c_1 = - c_3$ in \eqref{vecgenkin}. We again have four
independent fields, namely the four components of $A_a$.  In this
case, the canonical momentum $\Pi^0$ defined in \eqref{pi0def}
vanishes automatically, giving us a constraint:
\begin{equation} \label{A0constraint} \Pi_0 = 0 \equiv \Phi_1.
\end{equation}
The other three canonical momenta $\vec{\Pi}$ defined in
\eqref{pivecdef} have an invertible relationship with the
corresponding field velocities:
\begin{equation}
  \vec{\Pi} =
  c_1 \left( - \dot{\vec{ A}} + \vec{\nabla} A_0 \right). 
\end{equation}
Thus, the base Hamiltonian density $\mathcal{H}_B$, given by
\begin{equation}
  \mathcal{H}_B = \Pi_0 \dot{A}_0 + \vec{\Pi} \cdot \dot{\vec{A}} -
  \mathcal{L}
\end{equation}
must be augmented by an auxiliary Lagrange multiplier term enforcing
the constraint $\Pi^0 = 0$.  After simplification, this yields
\begin{multline}
  \mathcal{H}_A = -\frac{1}{2 c_1} \vec{\Pi}^2 + 
  \vec{\Pi} \cdot \vec{\nabla} A_0 \\ - \frac{c_1}{2} \left[ (\partial_i A_j)
  (\partial^i A^j) - (\partial_i A_j) (\partial^j A^i) \right] \\
  {}+
  \kappa (A^2 -b)^2 + u_0 \Pi^0.
\end{multline}
Once again, we must ensure that the primary constraint
$\Phi = \Pi_0 = 0$ is conserved by the equations of motion; this again
produces a series of secondary constraints:
\begin{align}
  \dot{\Pi}^0 &= \{ \Pi_0, H_A \} = \vec{\nabla} \cdot \vec{\Pi} +
                4 \kappa (A^2 -b) A_0 \equiv \Psi_1 \\
  \dot{\Psi}_1 &= \{ \Psi_1, H_A \} \notag \\
              &= -8 \kappa \left[ \vec{\nabla} \cdot ((A^2 - b)
                \vec{A}) - (-3A_0^2 + \vec{A}^2 -b) 
                u_0 \right. \notag \\
              & \qquad \qquad \qquad \left. {} + A_0 \vec{A} \cdot
                \left(\frac{1}{c_1} \vec{\Pi} - \vec{\nabla} A_0\right) \right]. 
\end{align}
So long as $- 3 A_0^2 + \vec{A}^2 -b \neq 0$, the demand that the
secondary constraint $\Psi_1$ be preserved by the evolution determines
the auxiliary Lagrange multiplier $u_0$ uniquely. We therefore have
four fields, two constraints (one primary, one secondary), and no
undetermined Lagrange multipliers; counting the degree of freedom
therefore yields
$$
N_{dof} = 4 - \frac{1}{2}(2) = 3.
$$

\subsubsection{Lagrange multiplier model \label{MaxwellLM}}

We now apply the same process to the vector model with a Lagrange
multiplier, \eqref{vectorLM}.  With the addition of the Lagrange
multiplier $\lambda$, we must also introduce a conjugate momentum
$\varpi$.  As in the scalar LM model, this vanishes identically,
yielding a second primary constraint:
\begin{equation}
\varpi = \frac{\partial \mathcal{L}}{\partial \dot{\lambda} }= 0 \equiv \Phi_2
\end{equation}
Including this constraint with an auxiliary Lagrange multiplier
$u_\lambda$ in the Hamiltonian density gives us the augmented
Hamiltonian density for the model:
\begin{multline}
  \mathcal{H}_A = -\frac{1}{2 c_1} \vec{\Pi}^2 + \vec{\Pi} \cdot
  \vec{\nabla} A_0 \\ - \frac{c_1}{2} \left[ (\partial_i A_j)
    (\partial^i A^j) - (\partial_i A_j) (\partial^j A^i) \right] \\ +
  \lambda (A^2 -b) + u_0 \Pi^0 + u_\lambda \varpi.
\end{multline}

We now derive the secondary constraints and see if their
time-evolution fixes the auxiliary Lagrange multipliers $u_0$ and
$u_\lambda$:
\begin{align}
  \dot{\Phi}_1 &= \{ \Pi_0, H_A \} = \vec{\nabla} \cdot
  \vec{\Pi} + 2 \lambda A_0 \equiv \Psi_1 \label{vecLMlambda} \\
  \dot{\Phi}_2 & =\{ \varpi, H_A \} = -A^2 + b \equiv \Psi_2 \\
  \dot{\Psi}_1 &=  \{ \Psi_1, H_A \} =
  - \vec{\nabla} \cdot \left( \lambda \vec{A} \right) + 2 A_0 u_\lambda +
  2 \lambda u_0 \label{vecLMdet1} \\
  \dot{\Psi}_2 &= \{ \Psi_2 , H_A \} = 2 A_0 u_0 - 2 \vec{A} \cdot
                 \left( \vec{\Pi} + \vec{\nabla} A_0 
                 \right) \label{vecLMdet2} 
\end{align} 
Assuming $A_0 \neq 0$, the requirement that both \eqref{vecLMdet1} and
\eqref{vecLMdet2} vanish determines the auxiliary Lagrange multipliers
$u_\lambda$ and $u_0$.  The degree of freedom counting is therefore
five fields (four components of $A_a$, plus $\lambda$); four
constraints (two primary, two secondary); and no undetermined
auxiliary Lagrange multipliers, for a result of
$$
N_{dof} = 5 - \frac{1}{2}(4) = 3.
$$

This is a surprising result: the number of degrees of freedom of the
theory when the vector field is ``constrained'' to a vacuum manifold
determined by $f(A_a) = 0$ is exactly the same as when it is
``allowed'' to leave this vacuum manifold.  In other words, the
Lagrange-multiplier ``constraint'' does not actually reduce the
degrees of freedom of the model.

We can again connect this model to the terminology of
\cite{Garfinkle2012}, as we did for the ``general'' LM case in Section
\ref{genvecLM.sec}.  In that work, for a model of a timelike vector
field with $c_1 \neq 0$ and $c_{13} = 0$ (i.e., $c_3/c_1 = -1$), the
number of constraints was found to be zero for all points in
configuration space; such a model was therefore ``safe'', with three
degrees of freedom at all points in field space.  This is in agreement
with our work here: so long as the vector $A_a$ is constrained to be
timelike ($b < 0$), we will always have $A_0 \neq 0$, and the above
analysis holds.

\subsection{$V$-field case:  $c_{13} \neq 0$, $c_1 = 0$ \label{Vvec.sec}}

\subsubsection{Potential model}

In this case, we have a Lagrange density with $c_1 = 0$ and $c_3 \neq
0$; such a field is called a ``$V$-field'' by Isenberg \& Nester
\cite{Isenberg1977}.  When we calculate the conjugate momenta in this
case, \eqref{pi0def} allows us to solve for the velocity $\dot{A}_0 =
\Pi^0/c_3$, but \eqref{pivecdef} becomes a set of three constraints:
\begin{equation}
  \vec{\Phi} \equiv \vec{\Pi} + c_3 \vnab A_0 = 0. \label{Vfieldconstr}
\end{equation}
The augmented Hamiltonian density for the potential model \eqref{vectorpot} is then
\begin{align}
  \mathcal{H}_A 
  &= \Pi^0 A_0 + \vec{\Pi} \cdot \dot{\vec{A}} -
    \mathcal{L}_P + \vec{u} \cdot \left( \vec{\Pi} + c_3 \vnab A_0
    \right) \notag \\
  &= \frac{1}{2 c_3} \left(\Pi^0\right)^2 - \frac{c_3}{2} (\partial_i
    A_j) (\partial^j A_i ) \notag \\ 
  & \qquad \qquad {} + \kappa (A^2 - b)^2 + \vec{u} \cdot
    \left( \vec{\Pi} + c_3 \vnab A_0 
    \right),
\end{align}
where $\vec{u}$ is a vector of auxiliary Lagrange multipliers
enforcing the primary constraints \eqref{Vfieldconstr}.  Enforcing
these primary constraints under time evolution then yields a set of
three secondary constraints $\vec{\Psi}$:
\begin{equation}
  \dot{\vec{\Phi}} = \{ \vec{\Phi}, H_A \} = \vnab \Pi^0 - c_3
  \vnab \left( \vnab \cdot \vec{A} \right) - 4 \kappa (A^2 - b)
  \vec{A} \equiv \vec{\Psi}.
\end{equation}
The time-evolution of these secondary constraints, written out in
terms of spatial components, is then
\begin{multline}
  \dot{\Psi}_i = \{ \Psi_i, H_A \} = \mathcal{M}_{ij} u_j - 4 \kappa \partial_i
  \left( (A^2 - b) A_0 \right) \\+ \frac{8 \kappa}{c_3} A_0 A_i \left( \Pi^0 + \vnab
    \cdot \vec{A} \right), \label{Vfieldueqn}
\end{multline}
where 
\begin{equation}
  \mathcal{M}_{ij} \equiv 4 \kappa \left[ \delta_{ij} \left( A^2 - b\right) + 2
      A_i A_j \right].
\end{equation}
We require that $\dot{\Psi}_i = 0$.  This can be guaranteed in
equation \eqref{Vfieldueqn} via an appropriate choice of $u_j$ so long
as the matrix $\mathcal{M}_{ij}$ is invertible.  For general field
values, this inverse can be calculated to be
\begin{equation}
\left(\mathcal{M}^{-1} \right)_{ij} = \frac{1}{4 \kappa (A^2 - b)}
\left[ \delta_{ij} - \frac{2 A_i A_j}{1 + 2 \vec{A}^2} \right]
\end{equation}
and so we can solve the equation $\dot{\vec{\Psi}} = 0$ for $\vec{u}$
so long as $A^2 - b \neq 0$.\footnote{It is notable that this set of
  field values is precisely the vacuum manifold.  This property
  becomes more important in the context of tensor models involving
  potentials and Lagrange multipliers, and will be discussed more
  extensively in an upcoming work \cite{tensorpaper2018}.}  The generic theory
therefore has four fields, six constraints (three primary, three
secondary) and
\begin{equation}
  N_{dof} = 4 - \frac{1}{2} (6) = 1
\end{equation}
degree of freedom.

\subsubsection{Lagrange multiplier model}

As in Section \ref{MaxwellLM}, the switch from a potential $V$-field
model to a Lagrange-multiplier $V$-field model does not actually
``kill off'' any degrees of freedom.  The augmented Hamiltonian
density now contains one more auxiliary Lagrange multiplier
$u_\lambda$, which (as before) enforces the constraint $\varpi = 0$:
\begin{multline}
  \mathcal{H}_A = \frac{1}{2 c_3} \left(\Pi^0\right)^2 - \frac{c_3}{2} (\partial_i
    A_j) (\partial^j A_i ) + \lambda (A^2 - b) \\+ \vec{u} \cdot
    \left( \vec{\Pi} + c_3 \vnab A_0 \right) + u_\lambda \varpi.
\end{multline}
We thus have four primary constraints, $\vec{\Phi} = 0$ (as defined in
\eqref{Vfieldconstr}) and $\varpi = 0$.  Requiring that
$\dot{\vec{\Phi}} = 0$ and $\dot{\varpi} = 0$ then yields four
secondary constraints, which I will denote by $\vec{\Psi}$ and $\Psi$:
\begin{align}
  \dot{\vec{\Phi}} &= \{ \vec{\Phi}, H_A \} = \vnab \Pi^0 -
  c_3 \vnab \left( \vnab \cdot \vec{A} \right) - 2 \lambda \vec{A}
  \equiv \vec{\Psi}, \\
  \dot{\varpi} &= \{ \varpi, H_A \} = A^2 - b \equiv \Psi.
\end{align}
The time-evolution of these secondary constraints is then
\begin{align}
  \dot{\vec{\Psi}} 
  &= \{ \vec{\Psi}, H_A \} = 2 \left[ \vnab
    (\lambda A_0) - u_\lambda \vec{A} + \lambda \vec{u}
    \right] \label{VfieldLMueqn1} \\
  \dot{\Phi} 
  &= \{ \Psi, H_A \} = \vec{A} \cdot \vec{u} - \frac{1}{c_3}
    A_0 \Pi^0. \label{VfieldLMueqn2}
\end{align}
These equations determine all four auxiliary Lagrange multipliers
$\vec{u}$ and $u_\lambda$ so long as $\vec{A} \neq 0$ and
$\lambda \neq 0$;  in this case, we have
\begin{equation}
  u_\lambda = \frac{1}{\vec{A}^2} \left[ \vec{A} \cdot \vnab \left(
      \lambda A_0 \right) + \frac{1}{c_3} \lambda A_0 \Pi^0 \right]
\end{equation}
and 
\begin{equation}
  \vec{u} = \frac{1}{\lambda} \left[ u_\lambda \vec{A} - \vnab \left(
      \lambda A_0 \right) \right].
\end{equation}
Thus, for generic initial data, we are done.  We have five fields,
four primary constraints, and four secondary constraints;  and so the
number of degrees of freedom is
\begin{equation}
  N_{dof} = 5 - \frac{1}{2} (8) = 1.
\end{equation}
As for the Maxwellian vector theory in Section \ref{MaxwellLM}, the
addition of a Lagrange multiplier to a $V$-field model does not
reduce its degrees of freedom.

This analysis is again in agreement with the work of Garfinkle,
Isenberg, \& Martin-Garcia \cite{Garfinkle2012}.  For a model with
$c_1 = 0$, they find that Einstein-aether theory contains additional
initial data constraints on the three dynamical fields $\vec{A}$, and
is therefore ``endangered''.  In the present work, we have confirmed
this result: this model does indeed contain fewer than three degrees
of freedom.\footnote{The cases $\vec{A} = 0$ and $\lambda = 0$ were
  excluded from the above analysis.  In this case, one would have to
  look at the time evolution of the quantities in
  \eqref{VfieldLMueqn1} and \eqref{VfieldLMueqn2}, generate one or
  more second-order secondary constraints, and attempt to solve these
  for the auxiliary Lagrange multipliers.  In any event, this would
  generate a model with no more than one degree of freedom (if the
  resulting model was even consistent at such points in configuration
  space.)}

\section{Discussion \label{disc.sec}}

\subsection{Generalization}

We have found that a field theory model in flat spacetime may or may
not ``lose'' a degree of freedom when a constraint is added to the
system via a Lagrange multiplier.  Specifically, scalar models
(Section \ref{scalar.sec}) and general vector models (Section
\ref{genvector.sec}) lose a degree of freedom when we replace a
potential with a Lagrange multiplier; but Maxwell-type and $V$-type
vector models (Sections \ref{Maxvec.sec} \& \ref{Vvec.sec},
respectively) retain the same number of degrees of freedom regardless
of whether the field values are governed by a potential or by a
Lagrange multiplier.

There is an obvious difference between these cases.  In those models
where there are no primary constraints in the potential model, a
Lagrange multiplier eliminates a degree of freedom.  In contrast, in
the models where the potential model does contain primary constraints,
the field theory retains the same number of degrees of freedom when a
constraint is imposed via a Lagrange multiplier.

The reason for this difference can be traced to a particular feature
of the models we have examined.  In those models containing primary
constraints, the conservation of the first-order secondary constraints
leads to an equation that determines the auxiliary Lagrange multiplier
$u_\lambda$ (Eqns.~\eqref{vecLMdet1} and \eqref{VfieldLMueqn1} for the
Maxwell-like and $V$-field models, respectively).  In those models
without primary constraints, $u_\lambda$ is only determined once we
require that higher-order secondary constraints (specifically, the
third-order secondary constraints in Eqs.~\eqref{scalar.thirdordereq}
and \eqref{genvec.thirdordereq}) be conserved.

To extend this to a general statement, we first note that the primary
constraints for a potential model and its corresponding Lagrange
multiplier model are simply related.  If the primary constraints for
the potential model are a set of $M$ functions $\{\Phi_1, \cdots,
\Phi_M\}$, then the primary constraints for the corresponding Lagrange
multiplier model will simply be $\{\Phi_1, \cdots, \Phi_M, \varpi\}$,
where $\varpi$ is the conjugate momentum to the Lagrange multiplier
$\lambda$.  Moreover, $\varpi$ will commute with all of the primary
constraints that derive from the potential model, since none of these
constraints depend on $\lambda$.

The augmented Hamiltonian density will then be the base Hamiltonian
density with terms added to impose the constraints:
\begin{equation}
  \mathcal{H}_A = \mathcal{H}_0 + u_I \Phi_I +
  u_\lambda \varpi.
\end{equation}
(Here and in what follows, repeated capitalized Roman indices are
summed from 1 to $M$.)  The first-order secondary constraint required
in order to maintain $\varpi = 0$ under time evolution will then be
\begin{equation}
\Psi_\lambda = \{ \varpi, H_A \} = - \frac{\delta H_A}{ \delta
  \lambda} = f(\psi^\alpha), \label{genLMsecconstraint}
\end{equation}
where $\psi^\alpha$ here stands for the collection of fields in the
model.  In addition, there will be a set of first-order secondary
constraints $\Psi_I$ ($I = 1, \dots, M$), each derived from the
requirement that $\dot{\Phi}_I = 0$;  these are given by
\begin{equation}
  \Psi_I = \{ \Phi_I, H_A
  \}. \label{genLMsecconstraint2} 
\end{equation}

Now consider the time-evolution of the first-order secondary
constraints.  The time derivative of $\dot{\Psi}_\lambda$ will be
independent of $u_\lambda$, though it will generally depend on the
other auxiliary Lagrange multipliers $u_I$:
\begin{equation}
  \dot{\Psi}_\lambda \supset
  u_J \{ \Psi_\lambda, \Phi_J\} + u_\lambda \{
  \Psi_\lambda, \varpi \} = u_J \{ f(\psi^\alpha),
  \Phi_J\}. \label{genfieldsecevol} 
\end{equation}
Note that $\{f(\psi^\alpha), \varpi \} = 0$ since $f(\psi^\alpha)$ is
independent of $\lambda$.  Meanwhile, the expression
$\dot{\Psi}_I$ (for arbitrary $I$) will contain terms of the
form
\begin{equation}
  \dot{\Psi}_I = \{ \Psi_I, H_A \} \supset u_J \{
  \Psi_I, \Phi_J \} + u_\lambda \{
  \Psi_I, \varpi \}, \label{genLMsecevol}
\end{equation}
Equations \eqref{genfieldsecevol} and \eqref{genLMsecevol} together
imply that if
\begin{equation}
\{ \Psi_I, \varpi \} = 0, \label{genpsivarpicomm}
\end{equation}
then the equations for conservation of the constraints ($\dot{\Psi}_I
= 0$ and $\dot{\Psi}_\lambda = 0$) do not contain $u_\lambda$, leaving
this auxiliary Lagrange multiplier undetermined at this stage.  If
this occurs, then we must proceed to find additional second- and
higher-order secondary constraints.  Since we have more than two
additional constraints, but only one additional degree of freedom from
$\lambda$ itself, we conclude that in such cases, the
Lagrange-multiplier model will have fewer degrees of freedom than the
potential model.\footnote{It is also conceivable that $u_\lambda$
  could remain undetermined even after the process of finding the
  constraints is completed.  This would also reduce the number of
  degrees of freedom in the final counting.}

This condition \eqref{genpsivarpicomm} can be greatly elucidated via
use of the Jacobi identity.  Specifically, we have
\begin{equation}
  \{ \{ \Phi_I, H_A \}, \varpi \} + \{ \{ H_A, \varpi \}, \Phi_I \} + \{
  \{ \varpi, \Phi_I\} , H_A \} = 0
\end{equation}
for any primary constraint $\Phi_I$.  Since $\varpi$ commutes with the
rest of these primary constraints, the last term automatically
vanishes; and applying \eqref{genLMsecconstraint} and
\eqref{genLMsecconstraint2} yields the equation
\begin{equation}
  \{ \Psi_I, \varpi \} = - \{ f(\psi^\alpha), \Phi_I \}.
\end{equation}
Thus, the equation \eqref{genLMsecevol} will leave $u_\lambda$
undetermined, and the Lagrange multiplier will reduce the degrees of
freedom of the model, so long as
\begin{equation}
\{ f(\psi^\alpha), \Phi_I \} = 0, \label{primsurfcommute}
\end{equation}
i.e., the vacuum manifold function $f(\psi^\alpha)$ commutes with all
the primary constraints.

\subsection{Lagrange-multiplier models in dynamical
  spacetimes \label{curved.sec}} 

The number of degrees of freedom of a field theory in flat spacetime
is not always simply related to the number of degrees of freedom it
possesses in a curved, dynamical spacetime.  It is well-known that
diffeomorphism-invariant field theories have primary constraints
corresponding to the non-dynamical nature of the lapse and shift
functions; when we pass to a dynamical spacetime, we both introduce
new fields (the ten metric components) as well as new
constraints.\footnote{See \cite{Wald1984, Poisson2004} for a detailed
  description of the Hamiltonian formulation of general relativity.}
Perhaps less well-known, but equally important, is that degrees of
freedom which are unphysical (gauge or constraint) in flat spacetime
can become ``activated'' in a minimally coupled curved-spacetime
theory \cite{Isenberg1977}.  This occurs due to the fact that the
covariant derivative of a tensor field (unlike that of a scalar)
depends on the derivatives of the metric.  The ``minimally coupled''
kinetic term for a tensor field therefore contains couplings between
the metric derivatives and the tensor field derivatives, which can
turn equations that were constraints or gauge degrees of freedom in
flat spacetime into dynamical equations in curved spacetime, and vice
versa.

In light of these facts, we might then ask how much of the above
analysis would carry over to dynamical spacetimes.  Given the critical
role played by the constraints in this analysis, it is natural to ask
whether a Lagrange-multiplier model in a dynamical curved spacetime
would lose any degrees of freedom relative to the corresponding
potential model in a dynamical curved spacetime.

The condition \eqref{primsurfcommute} sheds some light on this
question.  We know that if $u_\lambda$ remains undetermined when we
require conservation of the first-order secondary constraints, then we
will in general have to find higher-order secondary constraints,
leading to a reduction of the degrees of freedom of the theory
relative to the corresponding potential model.  This will occur when
the vacuum manifold function $f(\psi^\alpha)$ commutes with the
primary constraints of the theory.

Any diffeomorphism-invariant theory, when decomposed into 3+1 form,
will contain terms involving the lapse $N$ and shift $N^a$;  these are
related to the spacetime metric $g^{ab}$ and the induced spatial
metric $h^{ab}$ by
\begin{equation}
g^{ab} = h^{ab} - \frac{1}{N^2} (t^a - N^a)(t^b - N^b),
\end{equation}
where $t^a$ is the vector field we have chosen to correspond to ``time
flow'' in our decomposition.  We can then write down the
Einstein-Hilbert action in terms of this induced metric, the lapse,
and the shift.  As the lapse and shift can be arbitrarily specified,
the are effectively ``gauge quantities'' corresponding to
diffeomorphism invariance; thus, their time derivatives do not appear
in the Lagrange density of the theory when it is decomposed.  In the
Dirac-Bergmann formalism, there are therefore primary constraints on
the momenta conjugate to these quantities:
\begin{align}
  \Pi \equiv \frac{\partial L}{\partial \dot{N}} 
  &= 0, 
  & \Pi_a \equiv \frac{\partial L}{\partial \dot{N}^a} &= 0.
\end{align}
The question is then whether the vacuum manifold function
$f(\psi^\alpha)$ commutes with these primary constraints.  But this is
easy enough to see, since
\begin{align}
  \{ f(\psi^\alpha), \Pi \} 
  &= \frac{\delta f}{\delta N}, 
  & \{ f(\psi^\alpha), \Pi_a \} = \frac{\delta f}{\delta N^a} 
\end{align}

Thus, the question of whether the Lagrange multiplier reduces the
number of constraints is reduced to the question of whether the vacuum
manifold function depends on the lapse and shift.  In particular, for
a collection of scalar fields in curved spacetime (the
dynamical-spacetime analogue of Section \ref{scalar.sec}), the vacuum
manifold function will be independent of the metric, and so there is
no way for the lapse or shift functions to enter into it.  We would
therefore expect that a Lagrange-multiplier model containing $N$
scalars would have fewer than $N$ degrees of freedom attributable to
the scalars.\footnote{As there is no coupling between the kinetic
  terms of the scalar and the metric, it seems likely that there
  would also still be two degrees of freedom attributable to the
  metric itself.}

However, for a function of a vector field $A_a$, the norm of the
vector field $A_a$ will depend on the lapse and shift functions:
\begin{align}
  A_a A_b g^{ab} &= A_a h^{ab} A_b - \frac{\left((t^a - N^a) A_a
                   \right)^2}{N^2} \notag \\
                 & = A^\perp_a h^{ab} A^\perp_b - \frac{\left(A_t - N^a
                   A^\perp_a \right)^2}{N^2},
\end{align}
where $A_t = t^a A_a$ and $A^\perp_a = h_a {}^b A_b$.  Any function of
the spacetime norm of $A_a$ will therefore depend on the lapse and
shift, and so the vacuum manifold function will not commute with the
primary constraints of the theory.  Given the results stated above, it
seems unlikely that the Lagrange multiplier would reduce the number of
degrees of freedom of such a theory.

It is interesting to note that this coupling occurs even if the
flat-spacetime theory does not contain any primary constraints, as in
the general vector models described in Section \ref{genvector.sec}.
Since the conservation of the first-order secondary constraints
determines the auxiliary multiplier $u_\lambda$ in this case, rather
than giving rise to further constraints, one would conclude that the
number of degrees of freedom of a general vector theory in curved
spacetime would \emph{not} be reduced by the presence of a Lagrange
multiplier, in contrast to the situation in flat spacetime.  In fact,
this is confirmed by known results.  A model consisting of a vector
field in a curved spacetime with a ``generic'' kinetic term (as in
Section \ref{genvector.sec}) will contain two ``metric'' degrees of
freedom and three ``vector'' degrees of freedom, regardless of whether
the vector is forced to a non-zero expectation value by a Lagrange
multiplier \cite{Jacobson2004} or by a potential \cite{Isenberg1977,
  Bluhm2008}.

\subsection{Potential models in the low-energy limit}

In classical particle mechanics, it is common to think of a
constrained system in relation to an unconstrained system with a
potential energy.  In the limit where the potential energy becomes
infinitely strong, it can be shown that the dynamics of the
unconstrained system reduce to those of a system constrained to lie
only in the minimum of the potential \cite{Arnold1989}.  It is
therefore common, in the analysis of constrained systems, to simply
include one or more Lagrange multipliers that enforce the constraints.
In general, each Lagrange multiplier reduces the number of degrees of
freedom of the system by one.

One might think that this general picture could be carried over to
field theory.  In particular, a set of fields in a potential could be
thought of as possessing a certain number of massive modes
(corresponding to oscillations in field-space directions in which the
potential increases) and a certain number of massless modes
(oscillations in field-space directions in which the potential is
flat.)  One could then construct a low-energy effective field theory
in which the massive modes have ``frozen out'', reducing the number of
degrees of freedom of the model.  In this low-energy limit, one would
expect the fields to always lie in their vacuum manifold, effectively
being constrained there.  Hence, one would think that the
Lagrange-multiplier version of a potential theory would nicely
correspond to the low-energy behavior of the corresponding potential
theory.

The results of this work, however, show that the picture is not so
simple.  While this simple picture holds for scalar fields in flat
spacetime, it seems quite unlikely that the low-energy limit of a
Maxwell-type or $V$-type vector field in a potential would correspond
to a model with the same kinetic term but containing a Lagrange
multiplier.  One would expect the low-energy limit to have fewer
degrees of freedom than the full potential model; but in these cases,
the Lagrange-multiplier models and the corresponding potential models
have the \emph{same} number of degrees of freedom.  While the
low-energy limit of some such models has been investigated
\cite{Bluhm2008,Seifert2009,Seifert2010b}, the Lagrange-multiplier
models would necessarily have a different behavior.  

In fact, this feature was noted in \cite{Bluhm2008} in the context of
a vector field with a ``Maxwell'' kinetic term.  In Section IV.C of
that work, it was noted that the Lagrange-multiplier model only
corresponded to the low-energy (``infinite-mass'') limit of the
potential model if the Lagrange multiplier $\lambda$ was set to zero
by fiat.  However, for a generic solution $\lambda$ will not vanish;
the vanishing of the secondary constraint in Eq.\ \eqref{vecLMlambda}
requires that $\lambda = - \vec{\nabla} \cdot \vec{\Pi}/2 A_0$.  In
other words, one must restrict the class of solutions under
considerations---i.e., further reduce the number of degrees of
freedom---to obtain the low-energy limit of a potential model from the
corresponding Lagrange-multiplier model.  This work shows that this
lack of direct correspondence is a common feature of models in which
tensor fields take on a vacuum expectation value.

\appendix*

\section{Poisson brackets and functionals}

In calculating the time-evolution of a field quantity in Hamiltonian
field theory, one would like to take the Poisson bracket of a field
$\psi^\alpha(x)$ with the Hamiltonian $H$ to find the time-evolution
of the field at $x$:
\begin{equation}
  \dot{\psi}^\alpha(x) = \{ \psi^\alpha(x), H \}.
\end{equation}
However, one does have to be careful with this notation, as the
Poisson bracket is only rigorously defined on real-valued field
functionals, not on functions of space like $\psi^\alpha$.
Specifically, we have
\begin{equation}
  \{ G_1, G_2 \} \equiv \int d^3 z \left[ \frac{\delta G_1}{\delta
      \psi^\alpha(z)} \frac{\delta G_2}{\delta
      \pi^\alpha(z)} - \frac{\delta G_1}{\delta
      \pi^\alpha(z)} \frac{\delta G_2}{\delta
      \psi^\alpha(z)} \right], \label{rigorPoissondef}
\end{equation}
where $\pi^\alpha$ is the conjugate field momentum to $\psi^\alpha$
(and a summation over $\alpha$ is implied), and the functional
derivatives are implicitly defined via the relation
\begin{equation}
  \delta G = \int d^3z \left( \frac{\delta G}{\delta
      \psi^\alpha(z)} \right) \delta \psi^\alpha(z).
\end{equation}

To extend the definition \eqref{rigorPoissondef} of a Poisson bracket
to a local field quantity $f(\psi^\alpha(x), \nabla \psi^\alpha (x),
\dots)$ constructed from field quantities at a fixed point $x$, one
introduces the functional
\begin{equation}
  F_x \equiv \int d^3 y \left[ f(\psi^\alpha(y), \nabla \psi^\alpha (y),
    \dots) \delta^3(x - y)
  \right].
\end{equation}
The functional derivatives in \eqref{rigorPoissondef} then become
\begin{multline}
\frac{\delta F_x}{\delta \psi^\alpha(z)} = \frac{\partial f}{\partial
  \psi^\alpha} \delta^3(x - z) - \nabla_a \left[ \frac{\partial f}{\partial
  (\nabla_a \psi^\alpha)} \delta^3(x - z) \right] \\ + \dots
\end{multline}
and similarly for $\pi^\alpha$, where the ellipses stand for
higher-order derivatives of $\psi^\alpha$ (or $\pi^\alpha$), and the
partial derivatives of $f$ (and their gradients) are evaluated at the
point $z$.  Here and throughout, I will use partial derivatives
$\partial$ to denote the variation of a locally constructed field
quantity with respect to one of its arguments, while the $\delta$
notation will be reserved for functional derivatives.

Under this extension, the Poisson bracket of a local field quantity
$f(\psi^\alpha(x), \nabla \psi^\alpha (x), \dots)$ with the
Hamiltonian $H = \int \mathcal{H} d^3x$ is ``really'' the Poisson
bracket of the functional $F_x$ with $H$.  Restricting attention to
quantities that only depend on the fields $\psi^\alpha$ and
$\pi^\alpha$ and their first derivatives, this Poisson bracket is
\begin{widetext}
\begin{align}
  \frac{d}{dt} \left[ f(\psi^\alpha, \nabla \psi^\alpha, \pi^\alpha,
  \nabla \pi^\alpha ) \right] 
  &= \{ F_x, H \} \notag \\
  &= \int d^3 z \left[ \left( \frac{\partial f}{\partial
    \psi^\alpha} \delta^3(x - z) - \nabla_a \left[ \frac{\partial f}{\partial
    (\nabla_a \psi^\alpha)} \delta^3(x - z) \right] \right) \frac{\delta
    H}{\delta \pi^\alpha (z)} \right. \notag \\ 
  & \qquad \qquad \qquad {} \left. - \left( \frac{\partial f}{\partial
    \pi^\alpha} \delta^3(x - z) - \nabla_a \left[ \frac{\partial f}{\partial
    (\nabla_a \pi^\alpha)} \delta^3(x - z) \right] \right) \frac{\delta
    H}{\delta \psi^\alpha (z)}\right] \\
  &= \frac{\partial f}{\partial \psi^\alpha} \frac{\delta
    H}{\delta \pi^\alpha (x)} + \frac{\partial f}{\partial
    (\nabla_a \psi^\alpha)} \nabla_a \left( \frac{\delta
    H}{\delta \pi^\alpha (x)} \right) - \frac{\partial f}{\partial \pi^\alpha} \frac{\delta
    H}{\delta \psi^\alpha (x)} - \frac{\partial f}{\partial
    (\nabla_a \pi^\alpha)} \nabla_a \left( \frac{\delta
    H}{\delta \psi^\alpha (x)} \right),
\end{align}
\end{widetext}
where all the field quantities are now evaluated at $x$.  

Note that this definition implies that that time-evolution ``commutes''
with spatial derivatives when we take the Poisson bracket, as one
would expect.  For example, suppose that $f = \psi^\alpha$ and $H =
\int \mathcal{H} d^3x$, where $\mathcal{H}$ is locally constructed
from the fields.  Then we have
\begin{equation}
  \{ \psi^\alpha, H \} = \frac{\delta H}{\delta \pi^\alpha (x)} =
  \frac{\partial \mathcal{H}}{\partial \pi^\alpha},
\end{equation}
evaluated at $x$.  Meanwhile, if $f = \nabla_a \psi^\alpha$, we have
\begin{multline}
  \{ \nabla_a \psi^\alpha, H \} = \nabla_a \left(  \frac{\delta
      H}{\delta \pi^\alpha (x)} \right) = \nabla_a \left(
    \frac{\partial \mathcal{H}}{\partial \pi^\alpha} \right) \\ =\nabla_a
  \left( \{ \psi^\alpha, H \} \right). 
\end{multline}
This fact simplifies the calculation of the Poisson brackets
considerably.

This definition can be extended straightforwardly to quantities
depending on higher derivatives of $\psi^\alpha$ and $\pi^\alpha$, and
the above-mentioned commutativity extends to such cases as well.  It
can also be extended to the Poisson brackets of two local field
quantities $f(x)$ and $g(y)$ by defining functionals $F_x$ and $G_y$
and following the same procedure. In such cases, the resulting Poisson
bracket will contain a factor of $\delta^3(x-y)$.  However, in the
interests of clarity, we will elide these factors when we take the
Poisson bracket of two such quantities; in other words, we will take
it as understood that the first argument of such a Poisson bracket is
evaluated at $x$, the second at $y$, and that the result is multiplied
by $\delta^3(x-y)$ or its derivatives.

\begin{acknowledgments}

I would like to thank D.\ Garfinkle, T.\ Jacobson, and J.\ Tasson for
helpful discussion and correspondence on this subject.  I would also
like to thank Perimeter Institute for their support and hospitality
during the period over which the majority of this research was
conducted.

\end{acknowledgments}

\bibliographystyle{unsrt}
\bibliography{vector_tensor_dofs}

\end{document}